\documentclass[prl,nofootinbib,twocolumn ]{revtex4}

\usepackage{graphicx}
\usepackage{amssymb}
\usepackage{amsmath}
\usepackage{amsfonts}
\usepackage{epstopdf}
\usepackage{epsfig}
\usepackage{wrapfig}

\begin{document}

\title{Breaking information-thermodynamics link}
\author{Robert Alicki,}
\affiliation{Institute of Theoretical Physics and Astrophysics, University of
Gda\'nsk, Poland }

\begin{abstract}
The information-thermodynamics link is revisited, going back to the analysis of Szilard's engine. It is argued that instead of equivalence rather complementarity of physical entropy and information theoretical one is a correct concept. Famous Landauer's formula for a minimal cost of information processing is replaced by a new one which takes into account accuracy and stability of information encoding. Two recent experiments illustrating the information-energy conversion are critically discussed.
\end{abstract} 
\maketitle

The long-lasting discussion  initiated in 1929 by the Szilard's analysis\cite{Szilard}  of Maxwell demon \cite{Maxwell} and followed by Brillouin \cite{Brillouin:1956},  Landauer \cite{Landauer:1961},  Bennett \cite{Bennett:2003}, Sagawa \cite{Sagawa:2012}, and others led to a generally accepted conclusion that a bit of information is equivalent to $k_B\ln 2 $ of thermodynamical entropy and  $ k_B T\ln 2 $ of energy, where $T$ is the temperature of environment. The common believe in this link strongly influenced the recent developments in quantum information theory and thermodynamics of nanoscopic systems.  Recently, this information-thermodynamics link has been illustrated using  nanoscopic implementations of the Szilard's engine \cite{Toyabe, Lutz}. Here, I show, by pointing out the missing energy source, that the interpretation of those experiments is based on a misconception already present in the original Szilard's analysis. I argue that information is an abstract entity independent of the physical nature of a carrier, which differs from the thermodynamical entropy and should not be included in the physical energy-entropy balance. Instead of equivalence, complementarity of information and thermodynamical entropy is proposed. The famous Landauer's formula for the minimal cost of information processing is replaced by the new one, based on a quantum model of a switch \cite{Alicki:2013, Alicki:2014}. 
\par
One should also comment on the numerous "derivations" of the Landauer's principle based on quantum statistical physics including the most recent and rigorous ones \cite{Reeb, Jaksic}.
In all those derivations the initial product state $\rho\otimes \omega_B$ of the information bearing degrees of freedom (typically a $d$-level system) and the heat bath is assumed. Although this is a useful approximation in the context of the derivation of Master equations it can be misleading when the balance of entropy for the total system is relevant. Namely, as the heat bath is a large system composed of $N$ particles with the density of states per unit energy scaling like $e^N$, we can replace the initial state by a slightly perturbed one $\rho_{SB}(\epsilon)$. This new initial state can differ in norm by $\epsilon$ and in energy by, say, $\epsilon k_BT$ , but its von Neumann entropy can deviate from the product state entropy by  $\sim \epsilon N$. This means that, from the point of view of the total entropy balance product state is not generic and unstable. Actually, preparation of a strictly product state in a realistic environment can cost work which should be included in the total thermodynamical balance.
\par
Szilard proposed a model of an engine which consists of a box, containing only a single gas particle, in thermal contact with a heat bath, and a partition.
The partition can be inserted into the box, dividing it into two equal volumes, and can slide without friction along the box. 
To extract  $k_B T\ln 2$ of work in an isothermal process of gas expansion one connects up the partition to a pulley. Szilard assumes that in order  to realize work extraction it is necessary to know  ``which side the molecule is on'' what corresponds to one bit of information.
\par
In this picture energy used to insert the partition is negligible  while the subjective lack of knowledge is treated as a real thermodynamical entropy reduced by the measurement. To avoid the conflict with the Second Law of Thermodynamics it is assumed that the reduction of entropy is compensated by its increase in the environment due to dissipation of at least $k_BT\ln 2$ work invested in the feed-back protocol of work extraction. This idea initiated a never ending discussion  about the place where the external work must be invested: in measurement\cite{Brillouin:1956}, in resetting memory\cite{Landauer:1961,Bennett:2003}, or both \cite{Sagawa:2012}.
\par
However, the reasoning of above is incorrect. First of all the insertion of partition compresses the single particle gas and reduces its entropy from the initial value $S$ to $S' = S - k_B \ln 2$ and hence increases the free energy  of the gas from the value $F = U - TS$ to the value  $F' = U - TS' = F + k_B T\ln 2$.\\  
 \emph{The excess  free energy $\Delta F = k_B T\ln 2$, which can be transformed into useful work, is objective and does not depend on anybody's knowledge  ``of which side the molecule is on''}.\\
 Indeed,  on the same level of idealization, one can design procedures of extracting work without knowing the position of the particle (see Fig. 1(A)). This was already noticed  by Popper and Feyerabend \cite{Feyerabend} who used a different design and  aimed  to reject the idea that statistical mechanical entropy was a subjective quantity.
\begin{figure}[tb]
    \centering
    \includegraphics[width=0.35\textwidth,angle=270]{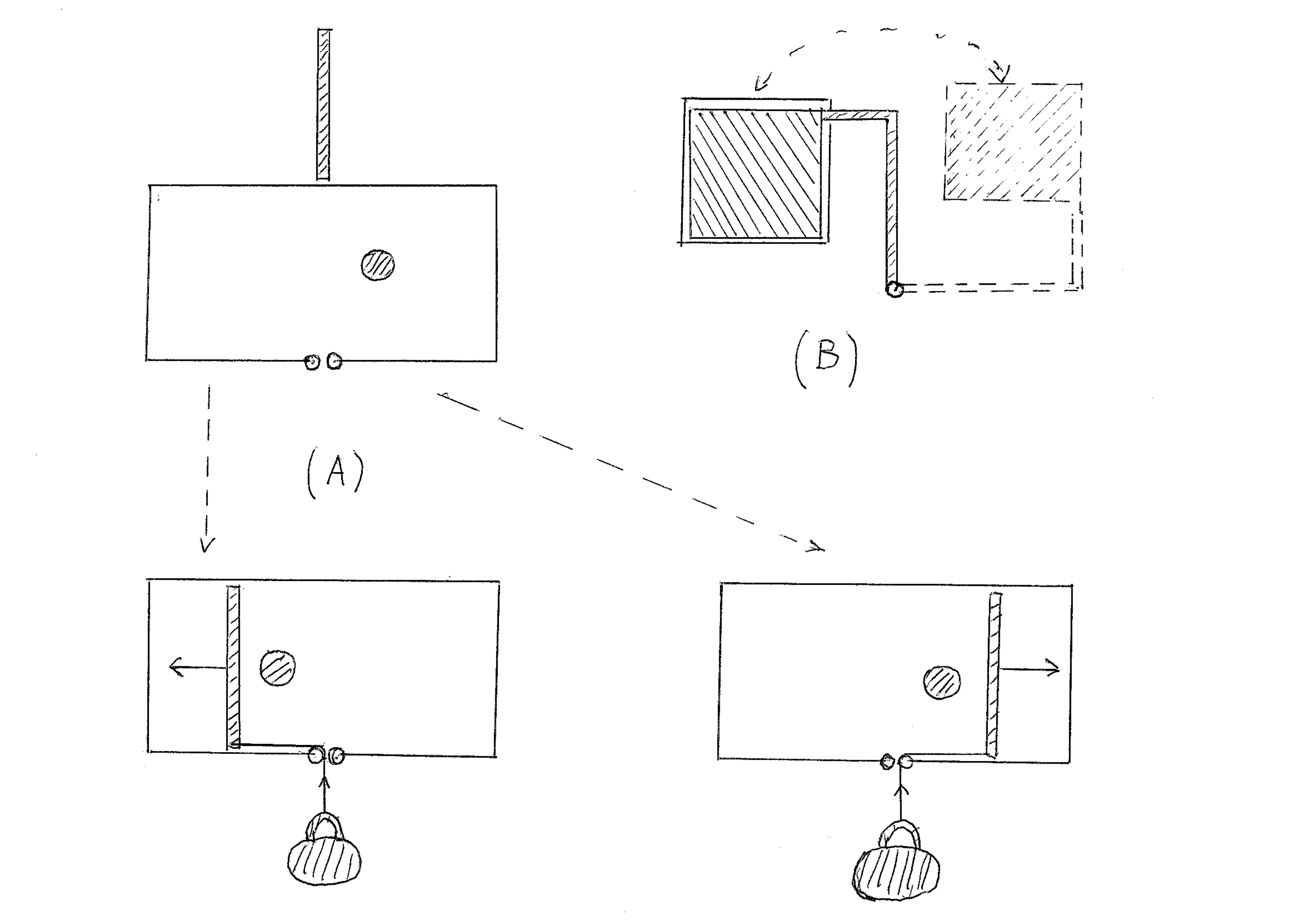}
    \caption{(A)  A design of the Szilard's engine which does not involve a measurement. (B) Partition as a switch with a potential barrier.}
    \label{nazwa1}
 \end{figure}
\par
The external work needed to reduce the physical entropy of gas particle is directly provided by the partition. Namely, the position of the partition inside the box must be stable with respect to thermal fluctuations and hence separated from the position ``partition outside the box'' by the energy barrier of the height $W >> k_BT$ (Fig.1(B)). Therefore, to move the partition from one position to the other one needs at least several $k_B T$ of work to overcome the energy barrier. The higher is the energy barrier the lower is the probability of error and hence the average extracted work is closer to the maximal value $k_B T \ln 2$.
Friction is necessary to stabilize the partition in the new position and the supplied work is partially dissipated into environment and partially converted into free energy of the gas particle. Summarizing:\\
\emph{ The operation of the Szilard engine can be easily explained without any reference to measurement and information}.\\

The statement above does not mean that feed-back control cannot be employed in the Szilard-type devices. It means only that it is superfluous and in principle can  be done at the arbitrarily low cost. The position of the particle in the box can be signalized by various types of interactions in a continuous fashion which cannot be quantified in terms of information measured in bits. Such a signal, conveyed with an infinitesimally low energy exchange, can trigger the mechanism driving the partition which is the real source of work.
\par
An example of such device has been realized  with the goal to present ``experimental demonstration of information-to-energy conversion''\cite{Toyabe}. A nanoscopic Brownian particle with a rotational degree of freedom can climb a spiral-staircase-like potential increasing its free energy, similarly to the Szilard's engine. Here, the ``partition''  is replaced by the controlled ``block'' which does not allow the particle to ``fall downstairs''. A random jump ``upstairs'' is observed and followed by putting the block one step higher, thus allowing only the increase of the free energy.
\par
The interpretation of this experiment  is based on the erroneous assumption that: ``in the ideal case, energy to place the block can be negligible''. In reality, ``placing a block'' (partition) costs at least several $k_BT$ of work and provides a direct energy injection into the system. The feed-back system which consists of a microscope and a digital camera  consumes an enormous amount of energy in $k_BT$ units per cycle and involves a lot of information processing but these costs are superfluous. The relevant energy input is provided by the electric field generator which implements the ``block''.
\par 
One of the consequences of  ``information-to-energy conversion'' paradigm is the claim that  $k_B T \ln 2$ is the minimal energy needed to execute an irreversible elementary logical gate. In particular this \emph{Landauer's formula} is applied to resetting an arbitrary logical value represented by a state of the information carrier to a fixed one represented by the given reference state. In fact the thermodynamical cost of information processing must depend on the accuracy and stability of information encoding: engraving on a stone tablet is more expensive than writing on paper. This means that the generally accepted Landauer's formula $W= k_BT \ln 2$ cannot give a practically useful bound for the cost of information processing. 
\par
To derive a more realistic bound for the cost of information processing I use a quantum model of a switch (see \cite{Alicki:2013, Alicki:2014} for technical details of this model). The switch  encodes a bit of information, and consists of two coupled quantum systems: a microscopic one, represented by spin-$1/2$, and  a harmonic oscillator  which represents the semi-classical pointer showing two macroscopically distinguishable positions of the switch. A kind of ``ferromagnetic coupling'' favorizes energetically particular position of oscillator (``right'' or ``left'')  correlated with spin polarizations (``up'' or ``down ``) ( Fig.2).

\begin{figure}[tb]
    \centering
    \includegraphics[width=0.35\textwidth,angle=90]{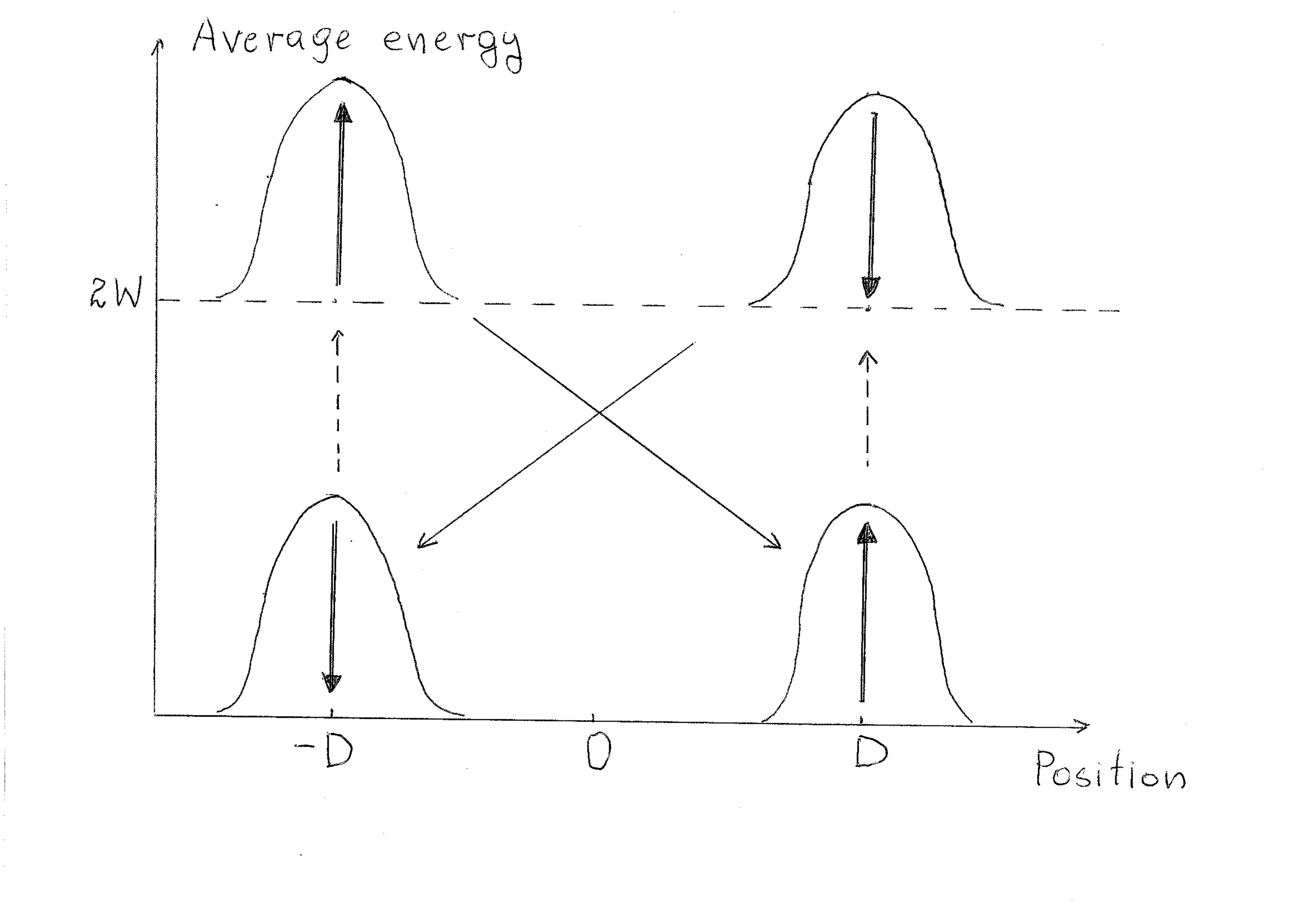}
    \caption{ Gaussians depict metastable localized oscillator states (bottom) or excited states (top) with arrows inside corresponding to spin direction. The long solid arrows represent dissipation routes, while the dashed ones spin-flip.}
    \label{nazwa}
 \end{figure}

The advantage of this model in comparison with  the common double-well potential one is that the former is exactly solvable even in the case of weak interaction with a heat bath. The picture which emerged from the detailed analysis  is the following. Two Gaussian mixed states of the quantum oscillator localized around two (dimensionless) positions $\pm D >> 1$ with the corresponding two spin states ``up/down'' are long living metastable states which encode a single bit of information. The reduced oscillator states describe switch positions which can be distinguished with the error probability  $\epsilon \in [0, 1/2]$. This error  is identified with the overlap of those states, called also \emph{quantum transition probability}  or \emph{fidelity}, and can be written as the Boltzmann-type factor
\begin{equation}
\epsilon =  \exp\Bigl\{-\frac{W}{\Theta}\Bigr\}  .
\label{highT}
\end{equation}
Here, $\Theta$ 
is the average quantum oscillator energy at the temperature $T$ interpolating between $k_B T$ (for $k_B T >> \hbar \omega_0$) and the oscillator zero-point energy $\hbar\omega_0/2$ (for $k_BT<< \hbar\omega_0$), thus characterizing both, thermal and quantum components of the environmental noise. The energy parameter $W$ is a half of energy amount necessary for spin-flip in one of the metastable states (Fig.2.) and can be interpreted as the ``energy barrier'' between the switch positions. Such a spin-flip caused by external time-dependent driving changes the switch position at the cost of $2W$ of work. This work is dissipated during the process of oscillator relaxation to the new position (Fig.2.) and is expressed as a function of error by the formula 
\begin{equation}
W = \Theta \ln\frac{1}{\epsilon}. 
\label{work}
\end{equation}
(compare with the Kish's formula $W= k_B T \ln (1/\epsilon)$ derived for the classical electronic switch \cite{Kish} and the same formula obtained by Brillouin in \cite{Brillouin:1956} for a particular example of measurement involving photons).
On average to reset the switch to the given reference position we need at least $W$  of work and hence  \eqref{work} should replace the Landauer's formula.
\par
The energy barrier protects the encoded information and this feature can be quantified by the life-time of the metastable states given by the Kramers-like formula
\begin{equation}
\tau = \tau_0  \exp\Bigl\{\frac{W}{\Theta}\Bigr\} = \tau_0 \frac{1}{\epsilon}. 
\label{kramers}
\end{equation}
Here, $\tau_0$ is the life-time of the spin  decoupled from the oscillator, but interacting with the same environment.
\par
The presented model of a switch is in a certain sense minimal and based on Gaussian quantum states what suggests that the obtained formulas \eqref{work} and  \eqref{kramers} provide universal relations for the minimal cost of resetting a bit and the proper scaling of the information life-time, respectively. One can expect that any attempt to ``recycle''
energy used to cross the barrier implies temporal reduction of the barrier and hence increase of error preserving the validity of the bound \eqref{work}.
\par
The relations \eqref{work},\eqref{kramers} imply much more realistic estimations for the thermodynamical cost of computation than the Landauer's formula. Again, such a cost should depend on the accuracy and stability of information processing. Hence, it should not be additive but the minimal cost of a single gate $W_{gate}$ must increase with the assumed total volume $N$ of the computation. It is not difficult to find such estimation in the limit of very large $N$ obtaining  $W_{gate} \simeq \Theta \ln N $ \cite{Alicki:2013, Alicki:2014}.
\par
Another ingenious experiment \cite{Lutz} demonstrates the process of resetting  of a single-bit memory implemented by the nanoscopic  Brownian particle in the controlled double-well potential. A simplified description of this process in terms of the Szilard setting is the following (Fig.3). Initially, the particle occupies the right or left part of the box and these two states are separated by the partition (potential barrier). One removes the partition and pushes the particle using a piston to the fixed, say right part of the box investing at least $k_B T\ln 2$ of work which is then dissipated into environment. Finally, one again inserts the partition. Exactly that dissipated work is measured in the experiment by tracking
the position of the Brownian particle and using the knowledge of the controlled time-dependent potential.\\ 
\emph{However, this amount of work bounded from below by $k_B T \ln 2$ is not the total cost of resetting}.
\par
Removing and inserting of the partition costs at least several $k_B T$ of work supplied by the external machinery which controls the shape of time-dependent potential with a high enough stability and accuracy. The related work bounded from below by the expression \eqref{work} with the known success rate $r= 1-\epsilon$ is not measured within this experimental scheme and for sure contains a large overhead. However, the reported maximal success rate $r = 0.96$ ($\epsilon = 0.04$) corresponds to the effective barrier height of $W = 3.2 k_BT$ predicted by the formula \eqref{work} and is consistent with the minimal ($2.2 k_BT$) and maximal ($\sim 8 k_BT $) values maintained during the experiment.

\begin{figure}[tb]
    \centering
    \includegraphics[width=0.35\textwidth,angle=90]{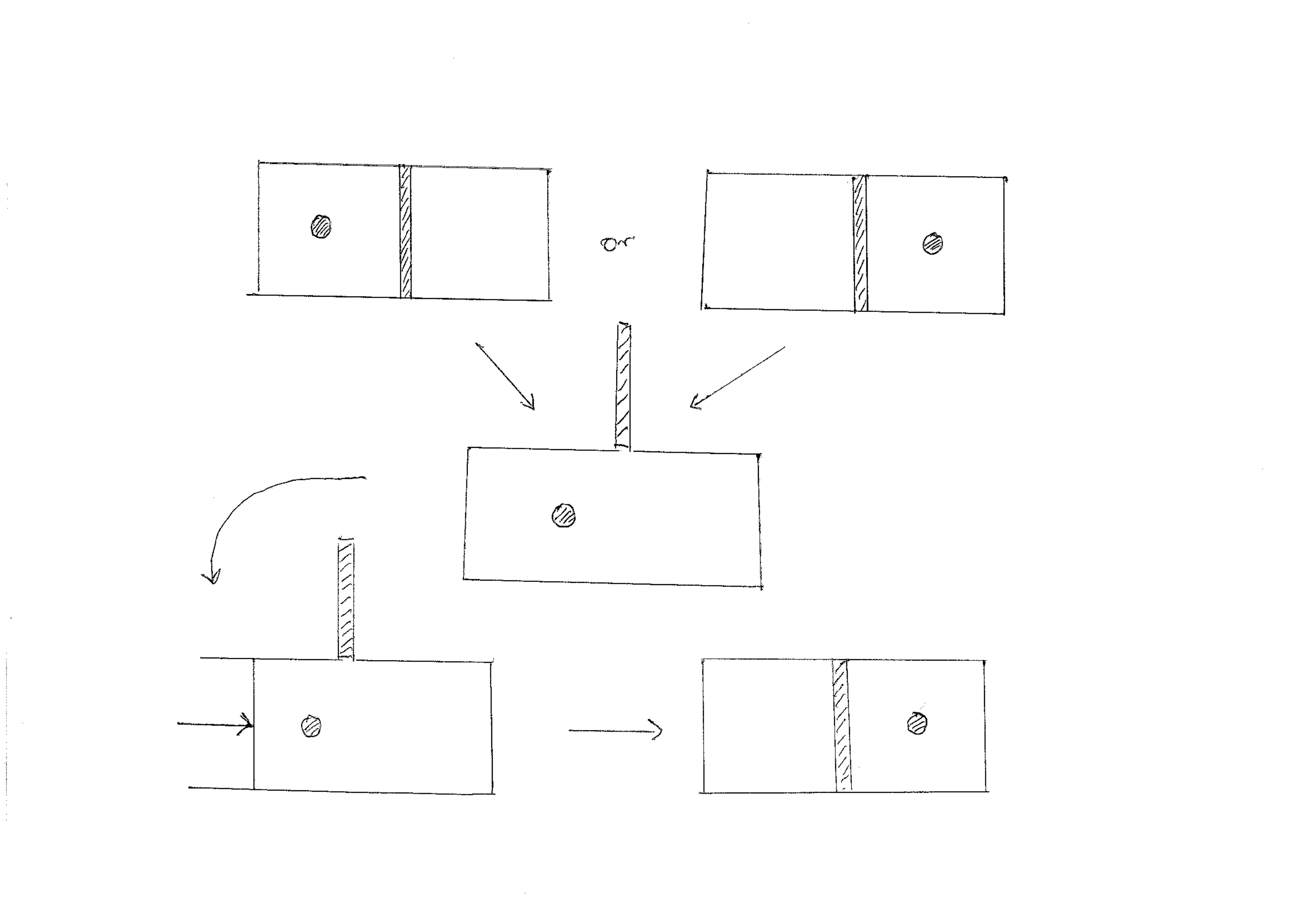}
    \caption{Resetting of a single bit encoded in the Szilard's machine}
    \label{nazwa3}
 \end{figure}

\par
The critical analysis of these two  experiments  and their theoretical background leads to the conclusion that their results do not contradict the statement that information is an abstract entity independent of the physical nature of a carrier. Information can be encoded in stable, long-living and well-distinguishable states of physical systems like, for example, ensembles of switches. The read-out of information can be done with an arbitrarily low cost, albeit with a given error. The real costs of information processing is the cost of any change of switch position, and must be at least as that given by the formula \eqref{work}. The subjective lack of knowledge about the position of a switch is not equivalent to the thermodynamical entropy and need not to be taken into account in the thermodynamical entropy-energy balance.
\par
Another instructive example of the fundamental difference between information and thermodynamical entropy is given by the comparison of two molecules: ammonia  ($NH_3$) and  alanine ($CH_3CH(NH_2)COOH$). For both molecules the model of a fictitious particle in a symmetric double-well potential gives the structure of lowest lying eigenstates. The ground state $\psi_0$ and the first excited state $\psi_1$ are symmetric and  antisymmetric superpositions of the states $\psi_L , \psi_R$ localized at the left or right potential minimum
\begin{equation}
\psi_0 = \frac{1}{\sqrt{2}}\bigl(\psi_L + \psi_R\bigr), \quad \psi_1 = \frac{1}{\sqrt{2}}\bigl(\psi_L - \psi_R\bigr).
\label{double_well}
\end{equation}
At the room temperature   equilibrium ensembles of both molecules should be described by the maximally mixed density matrix
\begin{equation}
{\rho}_{eq} = \frac{1}{2}\bigl[|\psi_0\rangle\langle \psi_0 | + |\psi_1\rangle\langle \psi_1 |\bigr]= \frac{1}{2}\bigl[|\psi_L\rangle\langle|\psi_L | + |\psi_R\rangle\langle \psi_R |\bigr]
\label{double_well1}
\end{equation}
with the von Neumann entropy equal to $S({\rho}_{eq} ) = \ln 2$.
\par
For ammonia the equilibrium distribution is maintained by fast relaxation processes caused by combination of collisional localization at $\psi_L$ or $\psi_R$  and tunneling with the inversion time of the order of $10^{-10} sec$. In ammonia maser a separator splits a beam of ammonia molecules and sends the molecules in excited state into the resonant cavity where coherent microwave radiation (work) is produced. Energy invested in the separation process  reduces the \emph{physical entropy} of the molecules increasing their free energy which is then  transformed into work. Hence, the von Neumann entropy of the density matrix  \eqref{double_well1} is the physical thermodynamical entropy. On the other hand, the unstable states of ammonia cannot be used as information carrier.
\par
For alanine both optical isomers $\psi_L ,\psi_R$ are extremely stable with the 
inversion time of the order of $10^{29}$ years. For example, in the biological context alanine appears always in the left-handed form. Therefore, only the states $\psi_L ,\psi_R$ of alanine can encode information in a very stable way and only the decomposition of the density matrix on the RHS of \eqref{double_well} seems to be physically meaningful. The von Neumann entropy represents here the \emph{subjective lack of knowledge} and has no thermodynamical  meaning. This follows from the fact that the absence of thermalization producing a mixture of left and right-handed states prohibits the work extraction.
\par
Therefore, instead of equivalence one can formulate \emph{complementarity principle for information and thermodynamical entropy} :\\
\emph{The von Neumann entropy of an ensemble of quantum states can be interpreted either as thermodynamical  or information-theoretical one depending on the stability of those states with respect to dynamics.}
\par
One usually assumes that all decompositions of a mixed states into pure or ``less mixed'' ones are equally meaningful. On the other hand, the so-called ``pointer states'' are examples of preferred basis \cite{Zurek:2003} which are structurally stable with respect to environmental noise. Hence, the role of dynamics  in the very definition of a mixed state (e.g. time-averaging vs. ensemble averaging) remains still worth of investigation.
\par
Although the validity and usefulness of the information-thermodynamics link has been criticized by several authors (using often arguments similar to those presented above\cite{Ishioka, Norton, Shenker}) a simple justification of these doubts in terms of  energy-entropy balance was missing. Here, the explicitly pointed out  source of work hidden in the operation of the Szilard's engine allows to avoid the controversial  information-theoretical considerations. This simple mechanistic approach gives an alternative interpretation of the experiments aiming to support the ``information-energy conversion''.
\par
Deconstruction of the information-thermodynamics link is quite important for the foundations of thermodynamics and related fields. The common believe in this link created the whole reality in which ``information'' reached the status of a ``substance''. This philosophy strongly influenced the development of \emph{quantum information}, \emph{quantum computing} and thermodynamics of nanoscopic systems.
\par
On the constructive side, the presented down to earth approach allows to derive more realistic bounds for energy dissipation per logical operation which take into account accuracy and stability of information processing. 
\par
The author acknowledges the support by the FNP TEAM project co-financed by EU Regional Development Fund.\\

\end{document}